\documentstyle[a4,epsf,rotate]{article}

\newcommand{\mathrm}[1]{{\rm #1}}
\newcommand{\qvev}{\langle\overline{q}q\rangle}
\newcommand{\ovpi}{\overline{\Pi}}

\begin{document}
\begin{titlepage}
\begin{flushright}
NORDITA-94/43 N,P
\end{flushright}
\vfill
\begin{center}
{\large\bf 2 and 3-point functions in the ENJL-model\footnote{presented by JB
at QCD94, July 7-13, 1994, Montpellier, France}.}\\
\vspace{2cm}
{J. Bijnens${}^a$ and J. Prades${a,b}^b$}\\[1cm]
{${}^a$ NORDITA, Blegdamsvej 17,\\
        DK-2100 Copenhagen \o, Denmark}\\[0.5cm]

 {${}^b$Niels Bohr Institute, Blegdamsvej 17,\\
        DK-2100 Copenhagen \o, Denmark}
\end{center}
\vfill
\begin{abstract}
We discuss the extended Nambu-Jona-Lasinio model as a low energy expansion,
all two-point functions and an example of a three-point function to all
orders in momenta and quark masses. The model is treated at leading level
in $1/N_c$ but otherwise exact. Some comments about the QCD flavour
anomaly and
Vector Meson Dominance in this class of models is made.
\end{abstract}
\vfill
August 1994
\end{titlepage}
% typeset front matter (including abstract)
\section{Introduction}

The extended Nambu-Jona-Lasinio (ENJL) model has phenomenologically been very
successful. For a recent review see \cite{Hatsuda}.
It is a typical example of a class of low energy hadronic models with quarks
that incorporates chiral symmetry correctly. The aim of this investigation was
twofold:\\
{\bf 1})
 to understand the relation between chiral symmetry and the concept
of constituent quarks.\\
{\bf 2})to understand why this class of models works so well
phenomenologically.\\
The main conclusions are:\\
{\bf 1}) in a sense the constituent quark mass
is the same as the quark-anti-quark vacuum expectation value.\\
{\bf 2}) The phenomenological success rests on 3 bases. The use of the $1/N_c$
expansion naturally leads to a kind of generalized meson dominance. The
short-distance behaviour of these models has a lot in common with QCD and a
large number of relations are fairly independent of the method of
regularization used.

The ENJL model has been treated for a long time and by many authors, see the
list of references in \cite{Hatsuda,BBR}. The Lagrangian is given by
\begin{eqnarray}
\label{LENJL}
%{\cal L}
\overline{q}\left[ i\gamma^\mu
\left(\partial_\mu - i v_\mu -i a_\mu \gamma_5 \right)
-{\cal M} - s + ip\gamma_5\right] q &&\nonumber\\
+ 2 g_S \sum_{i,j} (\overline{q}^i_R q^j_L )(\overline{q}^j_L q^i_R)
&&\nonumber \\
- g_V \sum_{i,j} |\overline{q}^i_L\gamma_\mu q^j_L |^2
+ |\overline{q}^i_R\gamma_\mu q^j_R|^2\ ,
&&
\end{eqnarray}
with $\overline{q} = (\overline{u}\ \overline{d}\ \overline{s})$ and
${\cal M} = \mathrm{diag}(m_u , m_d , m_s)$. $g_S$ and $g_V$ are given by
$g_{S(V)} = $ $4(8)\pi^2$ $G_{S(V)}/(N_c \Lambda_\chi^2)$
in terms of the notation used in \cite{BBR,BBZ,BP1,BP2}.
$v_\mu,a_\mu, s$and $p$
are external vector, axial-vector, scalar and pseudoscalar fields. These
are used to probe the theory with.
This Lagrangian can be argued to follow from QCD in the following way:
all indications are that in the pure glue sector there is a mass gap. The
lowest
glueball has  a mass of about 1.5~GeV. This means that correlations below
this scale should vanish. We then treat the interactions of quarks
below this scale as pointlike. It can also be seen as the first terms in an
expansion in local terms after integrating out fully (or only the short
distance part) of the gluons. These are in fact all terms up to dimension 6
to leading order in $1/N_c$. The Lagrangian in (\ref{LENJL}) also has the
correct chiral symmetry transformations.

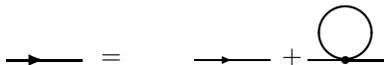
\begin{figure}[htb]
\unitlength 0.5cm
\begin{picture}(10,2)(-8,-0.5)
\thicklines
\put(-5,0){\vector(1,0){1}}
\put(-4,0){\line(1,0){1}}
\put(-2.5,-0.1){=}
\put(4,0){\line(1,0){1}}
\put(4,0.75){\circle{1.5}}
\put(4,0){\circle*{0.2}}
\thinlines
\put(3,0){\line(1,0){1}}
\put(2.3,-0.1){+}
\put(0,0){\vector(1,0){1}}
\put(1,0){\line(1,0){1}}
\end{picture}
\caption{The Schwinger Dyson equation for the propagator. A thin (thick) line
is the bare (full) fermion propagator.}
\label{figure1}
\end{figure}

\begin{figure}
\rotate[r]{\epsfysize=13.5cm\epsfxsize=8cm\epsfbox{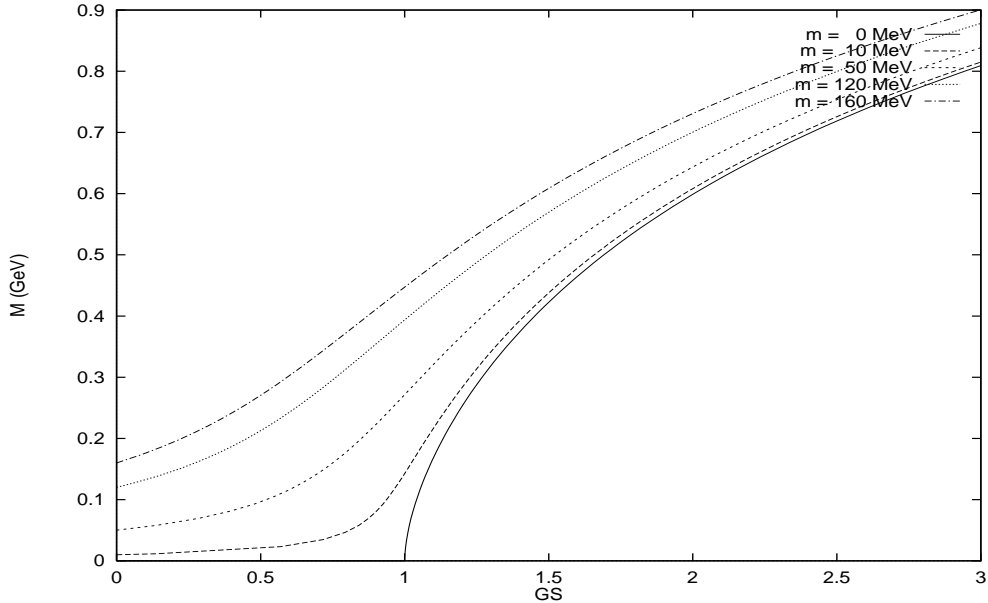}}
%\rotate[r]{\epsfysize=7.5cm\epsfxsize=5cm\epsfbox{fig2.ps}}
\caption{Plot of the dependence of the constituent quark mass $M_i$ as a
function of $G_S$ for several values of $m_i$}
\label{Figvev}
\end{figure}

Our first conclusion can be seen by looking at the Schwinger-Dyson equation
for the propagator depicted in Fig. \ref{figure1}.
This leads to the equation for the constituent quark mass ($M_i$) in terms of
the current one ($m_i$):
\begin{eqnarray}
\label{gap}
M_i &=& m_i - g_S \langle\overline{q}q\rangle_i\nonumber\\
\langle\overline{q}q\rangle_i &=& -N_c 4M_i \int\frac{d^4p}{(2\pi)^4}
\frac{i}{p^2 - M_i^2}\ .
\end{eqnarray}
The solution is shown in Fig. \ref{Figvev}. For nonzero current quark mass
there is only one solution. This one corresponds for $m_i\to 0$ to the
one with a nonzero value for $\langle\overline{q}q\rangle$ so that the chiral
symmetry is spontaneously broken. It also approaches this case smoothly
so chiral perturbation theory should be a valid expansion in this model.

\section{Low-energy expansion}

We use here the formulas of \cite{BBR}. We show here the results for the
low-energy parameters of chiral perturbation theory. Fitted are the
${\cal O}(p^4)$ parameters $L_i$ that are the free parameters at
next-to-leading
order. They are defined in \cite{BBR}. The scale is set by the only
dimensionful
parameter we fit here, the pion decay constant in the chiral limit, $f_0\approx
86$~MeV. We have redone the fits for a few new cases. One where we enforce
the one-gluon exchange relation $G_S = 4 G_V$, The one in the renormalon
picture
(Ref. \cite{Zakharov}) or $G_V = 0$ and with both $G_V$ and $G_S$ free.
These are column 3, 4 or 5 respectively in Table \ref{table1}.
Notice that the fit is significantly worse with $G_V = 0$. In \cite{BBR} this
fit was better due to the presence of a gluonic vacuum expectation value.
The value of the fitted
cut-off also decreases significantly with lower $G_V$ values.
\begin{table}
\begin{tabular}{ccccc}
   & exp &  & & \\
\hline
$G_V$ & - &$G_S/4$& 0 & 1.264\\
$M_Q$(MeV) & - & 260 & 280 & 265 \\
$\Lambda_\chi$(MeV)& - & 810 & 630 & 1160\\
\hline
$10^3 L_2$ & 1.2 & 1.5 & 1.6 & 1.6\\
$10^3L_3$&$-$3.6 & $-$3.1 & $-$3.0 & $-$4.1 \\
$10^3L_5$ & 1.4 & 2.1 & 1.9 & 1.5\\
$10^3L_8$ & 0.9 & 0.9 & 0.8 & 0.8\\
$10^3 L_9$& 6.9 & 5.7 & 5.2 & 6.7\\
$10^3L_{10}$&$-$5.5 & $-$3.9 & $-$2.6 & $-$5.5\\
$M_V$(MeV) & 770 & 1260 &$\infty$&810\\
$M_A$(MeV)&$\approx$1260 &2010&$\infty$&1330\\
\hline
\end{tabular}
\caption{Best fit values for the low energy parameters using
several constraints on $G_V$.}
\label{table1}
\end{table}
The value of $G_S$ is determined from $M_Q$ and $\Lambda_\chi$ via (\ref{gap}).

In addition to the good numerical fit a set of relations were obtained
that were independent of the regularization. Some of these are:
\begin{eqnarray}
L_9 &=& \frac{1}{2}f_V g_V\\
f_\pi^2 &=& f_V^2 m_V^2 - f_A^2 m_A^2 \ .
\end{eqnarray}
The former is the same as the vector meson dominance expression for the pion
vector form factor and the latter is the first Weinberg sum rule. The question
arises how general are these relations and are they broken at higher orders ?
Therefore we went on to analyze two-point functions to all orders in masses
and momenta.

\section{Two-point functions to all orders in $q^2$ and $m_q$.}

\begin{figure}
\begin{center}
%
% F:\TEX\TEXDRAW\ENJL2.TEX
% Mainfile for graphic-inclusion (created by TeX-Draw, JP-91)
%
%
\thicklines
\setlength{\unitlength}{0.5mm}
\begin{picture}(140.00,25.00)(0.,15.)
\put(97.50,35.00){\oval(15.00,10.00)}
\put(101.00,33.50){$\bigotimes$}
\put(17.50,35.00){\oval(15.00,10.00)}
\put(25.00,35.00){\circle*{2.00}}
\put(32.50,35.00){\oval(15.00,10.00)}
\put(40.00,35.00){\circle*{2.00}}
\put(47.50,35.00){\oval(15.00,10.00)}
\put(55.00,35.00){\circle*{2.00}}
\put(62.50,35.00){\oval(15.00,10.00)}
\put(06.00,33.50){$\bigotimes$}
\put(66.00,33.50){$\bigotimes$}
\put(86.00,33.50){$\bigotimes$}
\put(38.50,19.00){(a)}
\put(95.50,19.00){(b)}
\put(14.50,40.00){\vector(1,0){3.00}}
\put(29.50,40.00){\vector(1,0){3.50}}
\put(44.00,40.00){\vector(1,0){5.00}}
\put(60.50,40.00){\vector(1,0){3.00}}
\put(95.50,40.00){\vector(1,0){5.00}}
\put(99.00,30.00){\vector(-1,0){3.00}}
\put(64.00,30.00){\vector(-1,0){3.00}}
\put(49.50,30.00){\vector(-1,0){3.50}}
\put(34.00,30.00){\vector(-4,1){2.00}}
\put(18.00,30.00){\vector(-1,0){2.50}}
\end{picture}
\caption{The graphs contributing to the two point-functions
in the large $N_c$ limit.
a) The class of all strings of constituent quark loops.
The four-fermion vertices are those of \protect{\ref{LENJL}}.
The crosses at both ends are the insertion of the external sources.
b) The one-loop case.}
\label{Fig2pt}
\end{center}
\end{figure}
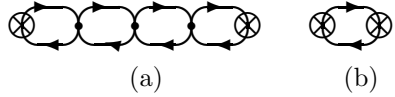

The chiral limit case was analyzed in \cite{BBZ}, the corrections due to
nonzero quark masses can be found in \cite{BP2}. Several relations were
found to be true to all orders. As an example we will derive here the relation
between the scalar mass and the constituent quark mass
in the chiral limit. The set of diagrams
that contributes is drawn in Fig. \ref{Fig2pt}a. The series can be rewritten
as a geometric series and can be easily summed in terms of the one-loop
2-point function $\overline{\Pi}_S$. The full result for the scalar-scalar
two-point function (we only treat the case with equal masses here, see
\cite{BP2} for the general case) is:
\begin{equation}
\Pi_S = \frac{\overline{\Pi}_S}{1- g_S \overline{\Pi}_S}\ .
\end{equation}
The resummation has generated a pole that corresponds to a scalar particle.
Can we say more already at this level?
We can in fact. The Ward identities for the one loop functions are:
\begin{eqnarray}
\label{WI1}
\ovpi_S &=& \ovpi_P - q^2\ovpi^{(0)}_A  \ ,\\
\label{WI2}
\ovpi_P &=& \frac{q^4}{4M_Q^2}\ovpi^{(0)}_A - \frac{\qvev}{M_Q}\ .
\end{eqnarray}
(\ref{WI1}) is a consequence of using the heat kernel for the one-loop
functions and (\ref{WI2}) is a direct consequence of the symmetry.
Using these two relations we can rewrite
\begin{equation}
1-g_S\ovpi_S = 1 + \frac{g_S}{M_Q} +
(q^2-4M_Q^2)\frac{q^2\ovpi^{(0)}_A}{4M_Q^2}
\ .
\end{equation}
The first two terms vanish due to the gap equation so this two-point function
has a pole at twice the constituent mass. For nonvanishing current
quark masses there is a small correction
\begin{equation}
M_S^2 = 4 M_Q^2 + g_A(-M_S^2) m_{ii}(-M_S^2)\ .
\end{equation}
See \cite{BP2} for definitions. Other examples of relations with the same
range of validity are:
\begin{enumerate}
\item The first and second Weinberg sum rule are satisfied, indicating a
somewhat too suppressed high energy behaviour for the last one.
\item The third Weinberg sum rule is violated as in QCD.
\item The Gell-Mann-Oakes-Renner relation to all orders in $m_q$ reads
$2 m_\pi^2(-q^2)f_\pi^2(-q^2) = (m_i+m_j)(M_i+M_j)/g_S$.
\end{enumerate}
These are valid in all schemes where the one-loop functions are obtained
from a heat kernel like expansion and have thus a rather broad range of
validity. In particular, they remain valid at finite temperatures and
densities.

\section{3 point functions and anomalies}

We discuss in this section as an example the pseudoscalar-vector-vector
3-point function. This has all the interesting features plus the occurrence
of the flavour
anomaly. The general diagram is a one-loop triangle diagram with
a chain  with 0,1,2,3,\ldots one-loop
(like in Fig. \ref{Fig2pt}a) connected to all three corners.
The two vector legs can be easily resummed leading to
an expression like:
\begin{equation}
\frac{g_{\mu\alpha} M_V^2(-p_1^2)-p_{1\mu}p_{1\alpha}}
{M_V^2(p_1^2)-p_1^2}
\end{equation}
The resummation of the external leg leads immediately to a VMD-like
formula in terms of slowly varying functions of the momenta. A similar
expression is valid for the other vector leg.
The pseudoscalar leg can mix with the longitudinal axial-vector degree
of freedom leading to a sum of two terms. Both with a pole at
the pseudoscalar mass. Naive use of the current identity
$-iq^\alpha\ \overline{q}\gamma_\alpha\gamma_5q =
2 M_Q i \overline{q}\gamma_5q$ would lead to only the pseudoscalar term
multiplied by $g_A(-q^2)$. It is in this way that this resummation method
sees the mixing of the pion and the axial-vector that occurs in this method.

In fact two more effects should be taken into account. The current identity
is used in a three-point function so there are usually also terms from the
equal-time commutators and there are additional terms in the current identity
due to the anomaly. These latter are very important in obtaining results that
have the correct QCD flavour anomaly \cite{BP1,BP2}.
The final result for the PVV two-point function with momenta $p_{1,2}$ for the
vector
legs is:
\begin{displaymath}
\Pi_{\mu\nu}^{PVV}(p_1,p_2) = \frac{N_c}{16\pi^2}
\frac{\varepsilon_{\mu\nu\beta\rho}p_1^\beta p_1^\rho 4 M_Q}
{g_S f_\pi^2(q^2)(m_\pi^2(q^2)-q^2)}
\end{displaymath}
\begin{displaymath}
\Bigg\{\frac{M_V^2(p_1^2) M_V^2(p_2^2)g_A(q^2)}{(M_V^2(p_1^2)-p_1^2)
(M_V^2(p_2^2)-p^2_2)} F(q^2,p_1^2,p_2^2)
\end{displaymath}
\begin{equation}
+ 1- g_A(q^2) \Bigg\}
\end{equation}
The function $F(q^2,p_1^2,p_2^2)$ is essentially the chiral quark loop result.
So analytically we only have a part that is multiplied by the expected
Vector-Meson-Dominance factors. There is a second part that is not, that
came from the extra terms in the current identity. This behaviour is in fact
very welcome. We have both the one-loop quark contribution to the slopes
and the one from vector meson dominance. Since both of these explain the
observed slopes having both fully would not agree with experiment.
Here we have, however
$F_{PVV}(m_\pi^1,p_1^2,p_2^2)$
$\approx$ $1 +$ $\rho(p_1^2+p_2^2)$ $+\rho^\prime m_\pi^2 +\cdots$
with $\rho
=g_A(0)\left(1/M_V^2(0)+ 1/(12M_Q^2)\right)$$\approx 1.53~\mathrm{GeV}^{-2}$
and
$\rho^\prime =$ $(g_A(0)/(12M_Q^2))\left(1-\Gamma(1)/\Gamma(0))
\right)$ $\approx$ $0.40$ $\mathrm{GeV}^{-2}$.
As we see we have good numerical agreement with the observed slope and the
corrections due to finite meson mass are substantially smaller than in the
chiral quark model. The latter is also desirable since otherwise there would
have been extremely large corrections to the $\eta$ decay.

\section{Meson Dominance}

As shown in the previous two sections the appearance of meson dominance
like formulas with slowly varying couplings is a natural feature
of this model and as such the successful phenomenology of this concept
is taken over. The model does combine this together with a set of
chiral quark loop effects in a kind of interpolating fashion thus
incorporating the strengths of both approaches.
The final results can be plotted to check whether the final formulas
also have a VMD-like behaviour and as shown in section 5 in \cite{BP2}
this is numerically the case for all the 2 and 3-point functions
studied there.
We have in general stayed in the euclidean domain of momenta to avoid the
problem that this model does not include confinement. There the sign of
meson dominance is that inverse formfactors are straight lines as a function
of the various $q^2$. This we find indeed.

\section{Conclusions}

We have treated the ENJL model in leading order in $1/N_c$ as an a low-energy
expansion for a general Green function via chiral perturbation theory
and all 2-point functions and a few 3-point functions to all orders
in quark masses and momenta. We have also given a consistent treatment
of the QCD flavour anomaly in this model.
So far the comparison with wanted QCD results and experiment has been very
successful.

\end{document}